# Practical IMT and EESS Spectrum Sharing in the 7 to 8 GHz Band[1]

Elliot Eichen (elliot.eichen@colorado.edu), Arvind Aradhya, and Oren Collaco
Computer Science, University of Colorado Boulder, Boulder CO USA

*Abstract—* The 7.3±0.175 GHz Earth Observation Satellite (EOS) band, while not protected, is used for Passive Sea Surface Temperature (P-SST) measurements that provide important data for weather forecasts, coastal disaster prevention, climate modeling, and oceanographic research. The full "7 GHz band" (7.125 to 8.4 GHz) – which encompasses these EOS frequencies – is the largest contiguous block of potentially available mid-band spectrum and will play a significant role in meeting the anticipated demand for wireless services. A Real-Time Geofenced Spectrum Sharing (RGSS) system is shown to be a practical and near-term solution to spectrum sharing between P-SST measurements and 5G/6G networks in the 7GHz band. RGSS enables IMT networks and EOS radiometers to share 350MHz of overlapping spectrum centered at 7.3 GHz. It prevents interference to P-SST measurements while simultaneously allowing IMT systems un-constrained access to the shared spectrum $\gtrsim$ 99.9% of the time. Subscriber impact during the $\lesssim$ 0.1% paused access time can be prevented by using 3GPP defined capabilities and O-RAN APIs to move subscribers to other frequencies. Paused access time data from a proof-of-concept RGSS system is available to academic, government, and industry researchers through a web or programmatic interface. *Keywords*—spectrum-sharing, geofencing, 5G, 6G, Sea Surface Temperature, EESS, 7-8 GHz

## I. INTRODUCTION

The "7 GHz" band (7.125 – 8.4 GHz) has been referred to as the 'Golden Band of [5 and] 6G." [1] It is the largest contiguous block of potentially available mid-band spectrum, and has similar coverage properties to 3.5GHz, enabling 7 GHz networks to reuse the footprint of base stations deployed for 3.5GHz thus expediting rapid deployment to meet the growing demand for wideband services [2]. The 7 GHz band is being evaluated as part of the US National Spectrum Strategy and by many ITU member states to understand the impact on incumbent applications – or, in some cases, the need to consolidate applications and/or move them to other frequencies. These evaluations will lead to discussions and potential recommendations at WRC-27 (Agenda Item 1.7) for making parts or all of the 7 GHz band available for wireless communications. The result of this process will lead to a heavily utilized, globally harmonized (i.e. with economies of scale) band that will shape the spectrum landscape for a generation.

Passive Sea Surface Temperature (P-SST) measurements provide important data for weather forecasting, climate modeling, coastal disaster prevention, oceanographic research, and "broad use in understanding changes to the marine and ecological environment" [4, 5]. Unfortunately, P-SST measurements by Earth Observation Satellites (EOS) at 6.925 and 7.3 GHz (the most important P-SST measurement bands because they provide good temperature sensitivity below 10°C) are not protected applications. Measurements at 6.925 GHz are now often corrupted in some coastal regions due to mobile networks, satellite communications, or point-to-point microwave transmissions [6, 7]. P-SST measurements centered at 7.3 GHz will be corrupted in coastal regions with dense IMT transmissions that overlap with (or are adjacent to) the measurement band. A spectrum sharing solution is required to protect P-SST measurements because the frequencies are fixed by molecular physics and cannot be moved. Imaging radiometers in the near IR (e.g. 3.6-4.1 μm and 10.1-12.2 μm [8]) currently provide some SST data but have difficulty penetrating cloud cover. New microwave bands (4.3 and 8.4 GHz [3]) have been proposed for P-SST measurement, as have onboard satellite signal processing methods to identify and remove radio frequency interference (RFI) [9, 10]. Both of these alternatives have advantages, but they also have drawbacks in terms of implementation time, cost, vulnerability to new spectrum allocations, and (in the case of on-board signal processing) high technology risk compared with the simplest solution of maintaining the availability of measurements at 7.3 GHz.

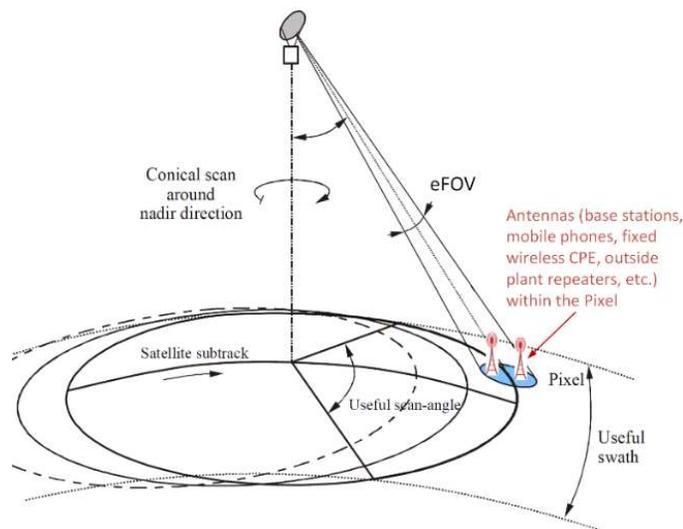

Figure 1: Conceptual view of RGSS (after [11] )

Another solution with the advantages of fast implementation time, flexibility, cost, and technological maturity is using geofencing to *prevent* RFI [12, 13, 14, 15]. Recent work has shown that Real-time Geofenced Spectrum Sharing (RGSS) can effectively protect EOS measurements at mm-wave frequencies while making this spectrum available for wireless service providers [16, 17] (Figure 1). Advantages of RGSS include:

- fast implementation time, and the ability to protect existing and new satellites carrying passive radiometers

- flexibility and speed in adapting to new network architectures, wireless devices, or more sensitive radiometers (inherent to a software solution),

- protection of radiometer measurements is based on *actual deployments*, rather than deployment models (which are often outdated and based on incorrect assumptions). This has significant advantages for both the wireless and the weather/climate communities.

- lower development and deployment costs (software for a band-agnostic system vs. band-specific hardware). \

- wide applicability to multiple bands. The methodology, software, and potentially the deployed system can be used at multiple frequencies. This helps with the "whack-a-mole" problem of assessing and preventing interference in new bands.

- simple and practical regulatory policing (base stations can be easily tested for compliance, and all transmitters can be programmatically audited) that also significantly reduces network element testing and design.

---

[1] This material is based in part upon work supported by the National Science Foundation under Grant No. 2232368.

The largest impediment to an RGSS implementation is that it requires cooperation between the wireless community (service providers, equipment vendors, etc.) and the weather/climate community.

This paper extends RGSS by evaluating the potential for IMT spectrum sharing with P-SST measurements at 7.3±0.175 GHz. The 7.3 GHz band is particularly amenable to real-time geofencing because only a few satellites are used for these measurements, thus relaxing constraints on the timing and enabling simple scheduling for IMT Radio Access Network (RAN) management. RGSS prevents interference to P-SST measurements while simultaneously making the entire measurement band available to IMT networks with less than 0.1% impact on network availability. Moreover, the impact on subscribers during the paused access time can potentially be mitigated or eliminated using existing Third Generation Partnership Project (3GPP) capabilities [18, 19, 20] and Open Radio Access Network (O-RAN) Application Program Interfaces (APIs) [21] that support frequency mobility.

## II. RGSS OPERATION IN THE 7 GHz EOS BAND

Conceptually, RGSS is a geofencing methodology. RGSS works by pausing transmissions for all network elements within the measurement footprint of a given scanning radiometer. There are no transmission restrictions on network elements outside of the measurement footprint. This is a reasonable premise because the "dark-time"/traversal – the pause time for any transmitter at a given position (lat/long) – is milliseconds to seconds, depending on the type of radiometer (cross-track or conical) and the operating parameters of the radiometer [16].

Calculating dark-times requires confidence that the geofencing boundaries are correctly calculated to eliminate interference. RGSS calculates the geofenced pixel size by doubling the radiometer's Field of View (FOV) where the FOV is defined as the Full Width at Half the Maximum (FWHM) value of the spatial power spectrum. This definition of a "geofenced" pixel – which is twice as big as the conventional definition of the measurement pixel – ensures that the maximum amount of RF leaking from outside the geofenced area is ≲ 2 % of the total power (68% of the energy is within the FWHM while 98% of the RF power is contained within twice the FWHM). In addition to doubling the radiometer's FOV, RGSS adds a one-pixel geographical guard band. For radiometers that do not synchronize their scan angle position to the satellite's orbital position (open loop systems), there is no way to predict the scan angle accurately at an arbitrary time based on actual position data downloaded from the satellite at a previous time. In this case, RGSS adds a full, non-overlapping scan line before the first dark-time scan line and after the last dark-time scan line.

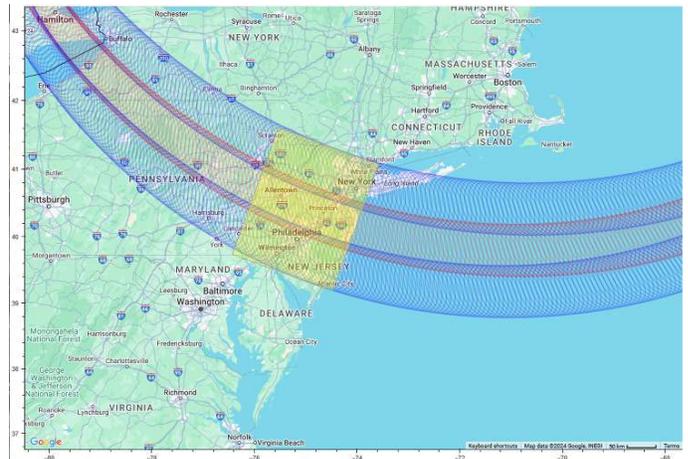

Figure 2: RGSS geofenced area and scan line swarths for a base station along the NJ shore (lat=42, long=-74) and AMSR2 orbital parameters @ 7.3 GHz on November 26, 2024 at ~7 UTC (2 AM local time). The light red swarth is the scan line that passes directly over the transmitter, while the light blue swarth are scan lines at beginning and end of the geofenced swarth. 14 scan lines (equal to 21 seconds) are required to geofence this traversal and the early afternoon traversal across the same coordinates. The yellow box represents the geofenced area (i.e. the number of impacted transmitters) located within 180 km from the coastal boundary (2 non-overlapping pixels).

Traversals from two satellites are needed to calculate the band availability for RGSS geofencing at 7.3 GHz: AMSR2 carried by JAXA GCOM2 and AMSR3 carried by JAXA GOSAT (which is planned for launch in 12/2024 [22]. AMSR2 and AMSR3 are conical scanning radiometers that make measurements in both horizontal and vertical polarizations at 11 frequencies, including the 6.925 and 7.3 bands used for P-SST measurements. To calculate the band availability, we assume that these radiometers run an open loop, and thus geofencing is done at the scan-line rather than pixel level. Using the 2FWHM definition of a geofenced pixel plus geographical guard band for AMSR2 requires 14 scans/traversal at 1.5 sec/scan for a total dark-time/day (2 traversals/day) of ~21 seconds. For two satellites, this yields a band availability of 99.96%. Even though the length of a radiometer length of a scan line for AMSR2 is ~2700 kms (122° full active scan angle at an altitude of ~710 km), the geofenced distance of 2 "geofenced" pixels is between ~90 and 180 km depending on the specific scan angle. From a carrier's perspective, the smaller the geofenced area, the fewer base stations that require RGSS to prevent interference with P-SST measurements.

As an example, Figure 2 shows the geofenced area and the top (blue), middle (red), and bottom (blue) scan line swarths for a transversal of AMSR2 over a base station close to the coastline near New York City.

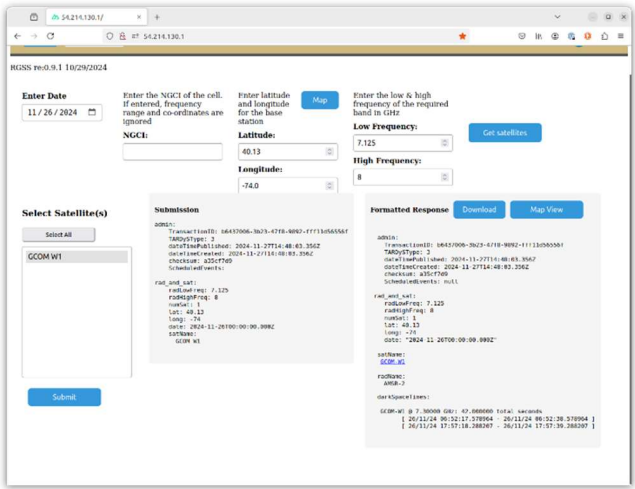

Figure 3 – RGSS Web Front End. A JSON API supports programmatic use. The system is available for use by academic, government, and industrial users.

Figure 3 shows the web front end of the RGSS POC (proof of concept) system. This front end is used for debugging and to visualize the geofencing output. Users provide a geographic location or the New Radio Cell Global Identity (NGCI is the unique code that identifies 5G network elements and that indexes that into a carrier's outside plant database) along with a date and frequency range. The system identifies all satellites that make passive microwave measurements within the given parameters, allowing users to select any or all of those satellites to calculate the required dark-times. The POC is programmatically available through a JSON API. The system currently supports NOAA 20 & 21 (ATMS), NASA SUOMI-NPP (ATMS), EUSAT (METOP B &C), and JAXA (AMSR2) satellites, and is available for use by academic, government, and industrial researchers. We intend to continue adding satellites and passive radiometers listed in the WMO OSCAR database.

While it is outside the main scope of this paper, it is worth noting that RGSS could also be applied to earth-to-satellite transmissions and point-to-point microwave links within the 6.925 P-SST GHz band. For IP-based layer 3 links, congestion and routing protocols could potentially be used to mitigate data loss during dark-time windows. In addition to AMSR radiometers, two new satellites carrying the Copernicus Imaging Microwave Radiometer (CIMR) will make P-SST measurements at 6.925 GHz beginning with EUSAT CIMR-A in 2029 and CIMR-B in 2031 [22].

III. COASTAL OPERATION

As discussed in section II, P-SST measurements at 7.3 GHz are vulnerable to interference from IMT transmissions in the 7 GHz band within ≈ 180 km (2 non-overlapping geofenced pixels) of the coastline[2]. This boundary is similar (but in some cases smaller) in area than the exclusion zones defined for Citizens Band Radio Service (CBRS) (Figure 3). (In the case of CBRS, exclusion zones are designed to limit interference to coastal radar systems defined by propagation models at 3.5 GHz.) 40% of the US population (and presumably at least 40% of 5G US wireless transmitters) reside within CBRS zones [23]. It is reasonable to suggest that there are similar parameters within and outside of the RGSS coastal boundaries and thus the same CBRS exclusion zones could be used to determine the number of network elements (gNBs, IABs, UEs, Fixed Wireless Access modems, etc.) inside an RGSS dynamic zone.

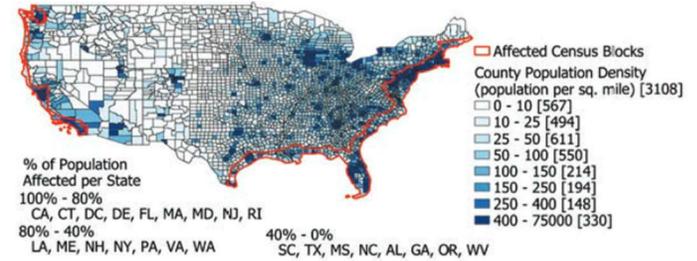

Figure 4. CBRS exclusion zones overlaid on population density (reproduced from reference [23])

IV. 5G/6G IMPLEMENTATION AND REGULATORY POLICING

Multiple architectures can be used to implement RGSS connected to a wireless service provider's (WISP) network. Figure 5 shows one such architecture that was designed based on feedback from colleagues at a major US WISP and an international equipment vendor. This architecture – which distributes functionality between a centralized swarth/pixel processor and in-network functionality to calculate dark-times and manage traffic – was designed to minimize the work needed to integrate with the WISP's network management system while keeping data from the WISP's outside plant database from being shared. (The outside plant database contains all network elements' locations and operating parameters. Sensitivity around this data is from the perspective of physical and cyber security, and maintaining the WISP's competitive advantage.). Colleagues at NOAA and NASA have suggested that a manual override of satellite orbital data be provided to accommodate changes in satellite orbits to avoid space debris. Some WISPs have proprietary requirements and interfaces to their network management systems and thus will require custom development.

---

[2] NOAA has suggested that for P-SST measurements, the coastline should also include the US Great Lakes [27].

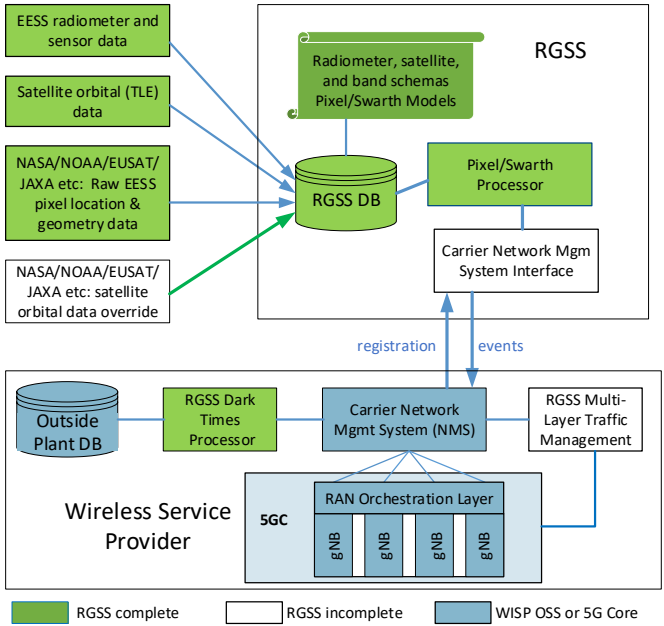

Figure 5: Distributed RGSS system architecture that enables wireless service providers to keep the details of their outside plant network confidential. Satellite orbital data and the raw EESS pixel location and geometry data are downloaded daily. The RGSS Multi-Layer Traffic Management modules would likely be provided to the 5GC (5G Core) system integrator.

From a system perspective, the major challenge for implementing RGSS is pausing transmission and moving wireless traffic to prevent any noticeable subscriber impact during the 21 seconds of dark-time for each satellite traversal. Fortunately, the 3rd Generation Partnership Project (3GPP, the wireless standards organization) has adopted mechanisms enabling 5G networks to move traffic to alternate frequencies to reduce network latency and system overhead by maintaining the radio link layer. These mechanisms include Dual Active Protocol Stack (DAPS) handover (3GPP rel 16, 2022) [19], and L1/L2-Triggered Mobility (3GPP rel 18, 2024) [20]. In addition to these mechanisms, network implementation could leverage session based Class of Service (CoS) identification to distinguish best-effort data streams from ultra-reliable low latency applications and real-time applications. A network implementation could also leverage WISP policies that sleep (idle) some base stations during periods of low traffic load (e.g., 1 AM to 5 AM local time). Traffic shaping in a mobile network would most likely be done by the multi-layer traffic management system using heuristic and machine learning algorithms.

To clarify the system implementation, consider GCOM-2 – the currently operational satellite used for P-SST measurements at 7.3 GHz – which makes early morning and mid day traversals (Figure 6).

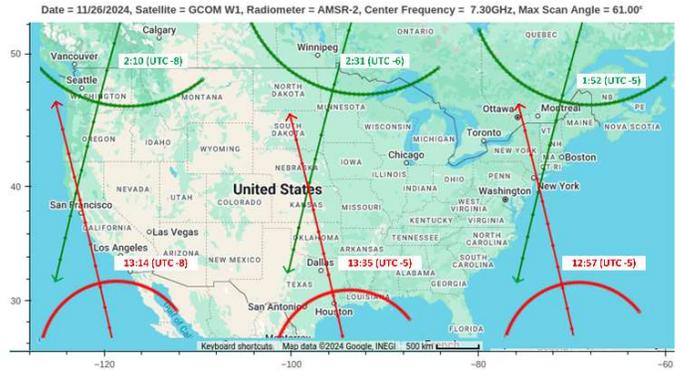

Figure 6: Traversals of AMSR2 on 11-26-2024. The UTC times of the Southbound (green) and Northbound (red) traversals have been replaced with local times.

Since the traffic load during the early morning traversals is often ≤15% of peak base station traffic, WISPs often put some base stations into a non-transmitting sleep mode to reduce power consumption (and thus operating costs). Putting gNBs (base stations) to sleep is a well-studied and common practice [24, 25] that RGSS could leverage to remove subscriber impact during the (21 second) dark-time window of the early morning traversal.

Moving traffic during the mid day dark-time traversal might be more difficult because this time of day has higher traffic levels. (Mobile traffic in urban US locations grows linearly from about 2 PM to a peak at about 9 PM local time, and then falls off to ~10-15% of peak traffic at about 4 AM). In this case, using the 5G 3GPP DAPS handover and/or L1/L2-Triggered Mobility capabilities could be employed to rapidly move additional wireless traffic to alternate bands during the dark-time window. Both DAPs and L1/L2 triggered mobility were designed to reduce the failure rate and the latency of legacy intercell handoff procedures [18]. DAPs uses a "make before break" paradigm to establish a second radio link to an alternate band gNB before triggering the handover. L1/L2 goes further by moving the transfer down the protocol stack to the medium access control (MAC) layer, which improves latency by pre-synchronizing the target data down and up links [18]. While both of these conditional handoff procedures were designed (in part) to help UEs initiate transfers in response to deteriorating physical layer performance, these handoff procedures also support network layer initiation as RGSS requires. Management of band transfers could use machine learning algorithms to prioritize sessions based on CoS markings. Additional work is required to demonstrate RGSS's ability to harness 5G's ability to move traffic quickly to alternate bands.

Finally, some members of the wireless community have suggested that zeroing specific physical resource block allocations could be used to shape the transmission frequency spectrum enough to limit interference within specific frequency limits [26]. Initial MATLAB simulations, however, indicate that the wide 350MHz P-SST measurement bandwidth and the residual noise floor of the blanked spectrum may make this option impractical for the RGSS application.

V. SUMMARY

The "7 GHz band" is the largest contiguous swarth of mid-band globally harmonized spectrum potentially available to meet the growing demand for wireless services. This will be an important topic of discussion and potential recommendation at the 2027 World Radio Congress. Passive sea surface temperature measurements - currently made in a 350 MHz window centered at 7.3 GHz - provide important data used for weather forecasting, climate modeling, coastal disaster prevention, and oceanographic research. These measurements, however, will be corrupted by 5G/6G transmissions in overlapping and adjacent frequencies.

This paper focuses on real-time geofencing, one potential solution to enable passive sea surface temperature measurements to share the 7.3 ± 0.175 GHz spectrum with 5G/6G wireless networks. The major results of this work are:

- a Real-Time Geospatial Spectrum Sharing (RGSS) system is a practical method to prevent interference with passive sea surface temperature measurements while simultaneously allowing 5G/6G networks access to the same spectrum greater than 99.9% of the time.
- 5G/6G network pause times ("dark-times") from a proof of concept Real-time Geospatial Spectrum Sharing system are currently available to academic, government, and industry researchers through a web interface or an application program interface.
- A system architecture that can centralize the real-time spectrum access functionality while preserving the confidentiality of carrier outside plant data (base station and other network element locations, operating parameters, traffic data, etc.) is presented.
- use of existing 3GPP specified inter-cell frequency movement capabilities and machine learning algorithms based on Class of Service (ultra-high reliability low latency applications, real-time applications, and best effort applications) are proposed to insulate network traffic from impact due to spectrum sharing.